\documentclass[%
preprint,
showpacs,preprintnumbers,
bibnotes,
 amsmath,amssymb,
 aps,
prstper,
longbibliography,
showkeys,
]{revtex4-1}

\usepackage{graphicx}
\usepackage{dcolumn}
\usepackage{bm}
\usepackage[english]{babel}

\usepackage{color}

\begin{document}

\title{A game-based educational method relying on student-generated questions}
\author{Enrique Abad}
\thanks{Author to whom the correspondence should be addressed (eabad@unex.es)}
\altaffiliation[also affiliated to ]{Instituto de Computaci\'on Cient\'{\i}fica Avanzada (ICCAEX), E-06071 Badajoz, Spain.}
\affiliation{
Departamento de F\'{\i}sica Aplicada, \\
Centro Universitario de M\'erida, Universidad de Extremadura, E-06800 M\'erida, Spain.\\
}
\author{Pilar Su\'arez}
\affiliation{Departamento de F\'{\i}sica Aplicada, \\
Escuela de Ingenier\'{\i}as Industriales, Universidad de Extremadura, E-06006 Badajoz, Spain.\\
}
\author{Julia Gil}
\affiliation{Departamento de F\'{\i}sica Aplicada, \\
Centro Universitario de M\'erida, Universidad de Extremadura, E-06800 M\'erida, Spain.\\
}

 \date{\today}

\begin{abstract}


Student-generated multiple-choice questions (MCQs) revised by the instructor were used to design an educational game and to set part of the exam in the framework of an elementary course in Photonics. An anonymous survey among the students revealed their general satisfaction with the method. The exam results of students who had been asked to author MCQs were compared to those of a control group subject to traditional teaching. The results of the former in the MCQ test of the exam were significantly better, without this being detrimental to their comparative performance in problem-solving. \\

\bf{Highlights}:

\begin{itemize}

\item Student-generated questions promote higher-order thinking.

\item The quality of student-generated questions often lies higher than expected.

\item Student-generated questions can be used for the design of activities stimulating the social interaction in the classroom and as a knowledge assessment tool.

\item An anonymous survey reveals the students' positive attitude towards the aforementioned uses of student-generated questions.

\item Games relying on student-generated questions might prove useful to enhance the students' motivation and to improve their academic performance.

\end{itemize}

\end{abstract}


\keywords{Student-generated questions, participatory learning, educational games, assessment tools.}
\maketitle


\section{\label{sec:lnt}Introduction}

Both for students and researchers it is important to provide answers to already existing questions; however, being able to properly formulate and to prioritize questions based on previous knowledge is of no lesser importance. The relevance of question posing in science is emphasized by a famous quote attributed to Einstein: ``If I had an hour to solve a problem and my life depended on the solution, I would spend the first 55 minutes determining the proper question to ask''.

By way of contrast, traditional teaching has largely failed to promote the students' question posing abilities. While traditional learning strategies based on instructor-generated questions, brain storming and drill-and-practice undoubtedly continue to be a must in the classroom, contemporary educational systems are increasingly placing emphasis on strategies assigning the student a more active role in learning and assessment processes \cite{Gil2010, Warren2010, Moreda2013}. Activities involving question-posing and answering are clearly one such strategy, the benefits of which have been widely documented in the existing literature on the subject.

Some of the above benefits are nicely summarized in ref. \onlinecite{Yu2009} as follows: ``developing a deeper understanding of the subject content learned, shifting from acquiring to using knowledge, achieving a sense of ownership of the subject contents as well as their learning experience, developing higher order thinking skills, generating more diverse and flexible thinking, becoming more involved in (and in control of) their learning, facilitating small group communication regarding the interacting topic, building up self-confidence about the subject matter and developing interest and ability in the follow-up problem-solving activity (...). In addition, problem-posing has value for the instructors of the classes as well, particularly because it reveals insight into students' abilities with the subject content and helps to provide an accurate assessment of what learners are capable of accomplishing (...)''.

Encouraging students to practice their oral skills by posing questions in the classroom is always a good idea. On the other hand, the composition of written questions provides a complementary and invaluable opportunity to develop a number of important cognitive and metacognitive skills, the proper use of academic language not being the least of them \cite{Bangert-Drowns2004}.  From the point of view of the instructor, it helps to evaluate the students' command of the relevant contents. Questions generated by students can be used to assess their understanding of theoretical principles and applications, their capability to communicate and to illustrate scientific ideas, and their ability to imagine situations where the acquired knowledge might play a role, e.g., real experiments or gedanken experiments.

Posing good questions and providing good answers requires the familiarity of the students with the course content and a deep previous reflection on how to formulate questions in a faultless and understandable manner. For a student with no previous experience, the first attempts might admittedly face considerable difficulty, requiring the instructor's guidance and feedback until the necessary level of training is achieved. Upon completion of this stage, question writing has proven to be a very useful basis to devise a number of training activities involving higher order skills.

As mentioned by Foos  \cite{Foos1989}, the recommendation of writing test questions as a learning method is occasionally found in books on how to succeed in college (e.g., ref \onlinecite{Lyng1986}), but only in recent years does this method appear to have been the object of systematic studies.  Horgen \cite{Horgen2007} was probably one of the first to carry out one such study.  In the framework of two IT courses, he let students write their own test questions and found clear learning benefits which he ascribed to an enhancement of the students' motivation. Even though the task was originally intended to be individual, some students spontaneously chose to collaborate with one another in the design of the written questions.

The presumably positive effects of collaborative strategies triggered the development of  PeerWise \cite{Denny2008}, a web-based tool allowing students to share their own multiple-choice questions (MCQs) and to receive feedback from their peers.  In this context, Yu and Liu pointed out the importance of anonymity as an element to create ``a psychologically safe learning environment'' \cite{Yu2009}. Other studies aimed at assessing the quality of student-generated MCQs find that it often lies above the instructors' expectations. This was for instance observed in an interesting study where the quality of MCQs authored by the students of two introductory Physics courses was comprehensively assessed \cite{Bates2014}. On the other hand, a recent work gives a word of caution against using student-generated questions as the sole source of on-line drill and practice materials, and suggests that better results might be achieved by combining them with instructor-generated questions \cite{Yu2014}.

Another recent study by Chang \emph{et al.} focused on the efficiency of embedding a game-based phase into a problem-posing scheme for mathematics learning in elementary education \cite{Chang2012}. In a similar spirit, here we use a game based on student-generated MCQs with the goal of rendering the students' preparation for the final exam more fun and interactive as well as improving their cognitive and metacognitive skills in crucial aspects.  Game-based methods and question writing have been successfully used to design formative and summative assessment tasks. Here, we go one step further and use a repository of student-generated questions to set the MCQ test for the final exam. While the use of games as educational tools appears to be more frequent in primary and secondary education, we strongly believe that it is also a suitable means to enhance the students' motivation at institutions of higher education. As far as games based on student-authored questions are concerned, we are not aware of any bibliographical references dealing with the subject in the context of higher education.

The present work was originally inspired by preliminary experiences carried out by one of us during the academic year 2009-2010. In the framework of an educational project, large groups of 1st year students prepared questions concerning various topics in electromagnetism. The student-generated questions were then reviewed by the instructor and the degree of difficulty of each question was rated. The questions were subsequently used to play ``Trivial Pursuit'' in the classroom, and the students were highly satisfied with the experience. Some time later, it dawned on us that quantitative studies could be carried out in order to assess the students' degree of satisfaction and possible learning benefits of games based on student-generated questions in a more systematic way.

A short description of our method is as follows: each student was asked to create her own list of MCQs on the course contents, comprising the question stem, one correct answer and three distractors. When the task was set, students were told that good questions could be used to set part of the final exam. The student-generated questions were then reviewed by the instructor to ensure that minimal quality criteria were met  \cite{Bates2014}. Unsuitable questions were discarded, suitable questions were retained, and potentially suitable questions were conveniently amended by the instructor. A repository of MCQs comprising the two latter categories and including the key to the correct answers was then made available to the students and subsequently used in a game consisting of a combination of  ``Trivial Pursuit'' and ``Millionaire'' -two of the most popular knowledge-based board games nowadays-.

The game was played on a Trivial Pursuit board and also followed similar rules; however, the structure of the repository questions used in the game formally resembled that of ``Millionaire'' (recall that each MCQ includes one correct answer and three distractors). A few weeks before the final exam, the game was played once in the classroom, and the students were also encouraged to subsequently meet and play the game by themselves as a preparation for the exam MCQ test. Finally, a subset of repository questions was selected by the instructor to set the MCQ test for the final exam. In that part of the exam, students were exclusively confronted with repository questions, that is, questions previously posed by their peers (in some cases conveniently amended by the instructor).  In practice, the large number of MCQs made it useless to learn the correct answers by heart. Thus, students were placed in a position somewhat analogous to examinees in driving license tests in Spain and other countries, who are required to provide the right answer for questions selected from a large repository. In view of the large number of questions, the advantage of knowing the possible questions in advance is strongly diluted, and there is little point in trying to memorize all the correct answers; instead, reasoning skills and practical experience prove to be much more effective as a learning strategy.

To carry out the present study, we chose the course ``Fundamentals of Photonics'', currently taught to 2nd year students in Telematics Engineering. Our study is based on the comparison of two groups of students: a control group subject to traditional teaching, and an experimental group of students who were required to author MCQs by themselves. The previous background of the students belonging to both groups and their academic performance in previous physics courses were very similar, thereby rendering a comparison of their results meaningful.

When analyzing the exam results of the control group and the experimental group, we found that the average score of the experimental group in the MCQ test was significantly higher, and their score in problem-solving was comparable with the control group's. By its own nature, our method requires a comprehensive revision by the instructor of a large number of student-generated questions. Thus, in the typical situation where limited human resources are available, the method can only be applied to small groups of students. Despite this statistical limitation and the fact that further tests are needed, the results of our statistical analysis are rather encouraging, and suggest that activities involving student-generated questions could be a suitable means to improve academic results, also in the context of higher education. An anonymous survey also reveals a positive attitude of the students towards the method.

The remainder of this work is organized as follows. Sec. \ref{sec:accont}  describes the educational context in which the method was used. Sec. \ref{sec:method} is devoted to a comprehensive description of our methodology. The results of the student poll and the statistical analysis are presented and discussed in Sec. \ref{sec:results}. Finally, in Sec. \ref{sec:conc} we give a summary of the main conclusions and outline possible avenues for future work.  The appendices are respectively devoted to an overview of the course program, examples of representative student-generated MCQs, the basic rules of ``Trivial Pursuit'',  and the survey questions answered by the students.

\section{\label{sec:accont} Educational context}

\subsection{Overview of the course program}

The course ``Fundamentals of Photonics'' is taught during the summer term of the 2nd year of Telematics engineering studies at the institution of two of us.  It provides an introduction to the properties of light and its interaction with matter, electromagnetic signal generation, transport, detection and conversion; it also deals with the working principles of some basic devices used in telecommunication and optoelectronic systems (e.g. lasers, laser diodes, LEDs, photodetectors, charge coupled devices, and photovoltaic cells). Comprehensive use of new technologies (e.g. java simulations) is made to support the explanations of key notions in the course.  The course comprises 7 chapters, whose contents are specified in Appendix A.

\subsection{Evaluation criteria}

The overall grade obtained by a student is a weighted average of the final exam grade (60\%), the grade obtained in follow-up activities organized by the instructor such as problem solving in the classroom (20\%), lab work (15\%) and oral participation in the classroom (5\%).  These criteria follow the guidelines and recommendations of the Spanish Agency for Quality Assessment (ANECA). The latter is responsible for the implementation of the Spanish national qualifications framework at higher education level (MECES) under the umbrella of the overarching qualifications network adopted by the 2005 Bergen conference of European Ministers responsible for Higher Education.

As already mentioned in the introduction, the final exam consists of two written parts, namely: a MCQ test (40\% of the exam score) and a problem-solving part comprising three problems (60\% of the exam score).

\section{\label{sec:method} Methodology}

The control group was formed by 13 students that attended the course during the summer term of the academic year 2011-2012. With this group, traditional teaching was used, involving classroom explanations, guided and autonomous problem-solving as well as guided lab work. Students were occasionally asked questions by the instructor in the classroom and also asked themselves questions spontaneously.

The remainder of this section describes the additional MCQ methodology used for an experimental group formed by 11 students during the summer term of the academic year 2012-2013. In what follows, the number of students is denoted by $N$. Thus, $N=13$ for the control group and $N=11$ for the experimental group.

\subsection{Preparation of MCQs}

At the beginning of the course, the students were given detailed information on the evaluation criteria and the weight of the different assessment tasks. They were also informed on the structure of the final exam, consisting of a MCQ test and three problems, as well as on the weight of each part in the exam score (40\%-60\%). The instructor also announced that in due time the students would be required to prepare MCQs on the course content, and that the latter would subsequently be used in a game resembling "Millionaire". The possibility that a subset of questions could be used to set the MCQ test for the final exam (if necessary with amendments performed by the instructor) was also explicitly mentioned.

Having explained the course chapters 1, 2 and 3, the instructor asked each student to provide 10 MCQs per chapter within three weeks. For every MCQ, each student had to provide an answer which she thought to be correct, as well as three additional distractors. Students were asked to author questions covering the topics of each chapter as uniformly as possible. They were also asked to provide questions of different nature focusing on theoretical principles and definitions, short calculations, problems solved in the classroom, experiments explained by the instructor or performed by the students in the lab, as well as gedanken experiments.

Students were explicitly asked to avoid trivial or flippant questions/distractors and warned that any attempt to plagiarize questions would be detected by appropriate means and could result in severe penalties. In any case, the questions posed should fall within the range of difficulty set by the instructor in the course. Students were given some examples of admissible questions corresponding to different levels of difficulty and asked to provide a miscellanea of questions with different degrees of complexity. Having said this, the instructor encouraged the students to think of interesting and original questions to enforce what had been learned in the classroom. However, students were asked to make use of their good judgment and to refrain from designing questions that could not be answered by resorting solely to what had been taught in the course.

Upon completion, the MCQs were uploaded by the students to our virtual campus via a Moodle interface. The questions uploaded by each student were subsequently reviewed by the instructor, who checked systematically that there was no significant overlap with questions uploaded by other students. In cases where such an overlap was detected, those students who had not been the first to upload the questions were requested to provide substitute MCQs.

A similar procedure was repeated with the course chapters 4, 5 and 6, whereby only 5 MCQs on the contents of Chapter 6 were required (this chapter deals with laser physics and is significantly shorter than the rest). This resulted in each student providing 25 questions on the contents of the aforementioned chapters. Accordingly, 10 participating students provided a total of 550 questions on chapters 1 to 6. One student chose not to participate in the preparation of MCQs, but he did participate in the game, and he also took the final exam.

In a second stage, the instructor comprehensively assessed the suitability of the MCQs. Out of the 550 questions prepared by the students, 195 were discarded, the final size of the repository thus became equal to 355. Some MCQs were discarded because of an unsuitable level of difficulty (even though a broad range of difficulty was admissible, some questions were too easy or even trivial; in exceptional cases, and despite previous warnings, the proposed questions were too difficult and clearly went beyond the level of difficulty set by the instructor in the course). In other cases, MCQs were discarded because they were ill-posed, too ambiguous, or because the language was too poor or too misleading, which would have made a full rewriting of the original question necessary. In very few cases, MCQs were also rejected because they did not correspond (partly or in full) to the taught contents.

In a non-negligible number of cases, original MCQs were made suitable and approved by the instructor upon introducing minor to moderate amendments and corrections. The questions were then delivered back to the students, who were given the opportunity to enquire at individual tutorials as to why some of their MCQs had been discarded or amended. Finally, the instructor granted to all the students access to the repository of 355 revised MCQs and the key to the correct answers. Students were told that these were the questions to be used in the game and they were given about three weeks to prepare for the latter.

Additionally, each student was asked to prepare two MCQs for the last course chapter. Due to time constraints, these MCQs were not used for the game. However, 16 out of 20 questions on the chapter contents were retained and incorporated to the repository with the key to the correct answers, giving a total of 371 MCQs that could potentially be used by the instructor to set the exam test.

The questions written by the students indeed turned out to be very diverse in nature and could be classified in different taxonomic levels of knowledge \cite{Bloom1956, Anderson2001, Bates2014}.  They referred to topics explained by the instructor, to problems that had already been solved in the classroom, and even to lab work. In many cases the students were able to author rather interesting questions.  Some representative examples of original MCQs authored by the students can be found in Appendix B.

\subsection{Playing the game}

The game strongly ressembles Trivial Pursuit, and it is played by very similar rules (see Appendix C), but repository MCQs rather than instructor-generated questions are used. The game was played on a Trivial Pursuit board, each color of the spaces in the wheel-like track corresponding to a different course chapter. For each chapter, a (numbered) list of MCQs was available, from which questions should be drawn. In order to save time, the game was not played individually, but rather by teams competing against one another (our 11 students thus corresponded to 5 different teams, four teams of two students and one of three students).  We also introduced a limitation in the number of questions in a row (5) that could be answered by the same team correctly without passing the turn to the next team, the aim being to avoid that strong teams monopolise the game and that enough time was granted to every team to effectively take part in it.

After an initial draw to set the starting order, the starting team would roll a die and move their playing piece along the board by the corresponding number of spaces. The color of the space on which the piece would land determined the list from which the MCQ to be answered should be drawn.  When being asked for the very first time in the game, each team were allowed to set their initial condition by choosing which of the team members should attempt to give the correct answer. The designated team member then had to identify the correct answer out of the four options provided (team mates were not allowed to communicate with one another during the game).

Once a student had given a correct answer to the question at hand, the turn was passed by rotation to the next team mate, and so the team were allowed to answer correctly up to five times in a row. As soon as a wrong answer was given or the maximum number of five correct answers in a row was reached, the turn was passed by rotation to the next team.  After one MCQ had been drawn and answered, it was eliminated from the corresponding list to avoid repetition in subsequent draws. The draw was carried out by rolling a virtual die (mimicked with a pseudorandom number generator) with as many faces as the number of remaining MCQs in the list of questions for each chapter. Very occasionally, it could happen that a student was lucky enough to be led to answer her own question, either in original form or with amendments made by the instructor.

Whenever the playing piece would land on category headquarter spaces at the end of each spoke of the wheel-like track, the team at hand could collect wedges of different colors by correctly answering the corresponding MCQs.  The first team to collect all six wedges, to reach the central hexagonal hub and to correctly answer a MCQ corresponding to a chapter of their choice would be declared the winner of the game. However, since only two hours were available to play the game, the instructor agreed with the students that if no team had been able to reach the central hub and to provide the correct answer to the final question within the specified time limit, the game would be won by the team that had the largest number of wedges in their playing piece. In case of a tie between two or more teams (equal number of collected wedges), the team having provided the largest number of correct answers during the game would be proclaimed winner.

Finally, the winning team were given a symbolic present (a glow lamp with a color changing RGB LED system) during a short award-giving ceremony. However, the winning team were explicitly warned that the MCQs chosen by the instructor for the final exam could be very different from the ones they had had to answer in the game.

\subsection{Final exam}

After the students had played the game and uploaded their MCQs on Chapter 7, it was clear that the repository questions were appropriate and numerous enough to be used as the source of MCQs for the exam test, and so the students were told that the exam test would contain repository MCQs only. However, they were not told whether there would be exam questions from each chapter or not.

About four weeks after the MCQs for the last chapter had been reviewed by the instructor and made available to the students, they took the final exam, consisting of a MCQ test with 20 questions selected from the repository (40\% of the exam score) and three problems to be solved (60\% of the exam score). The instructor chose the MCQs for the test according to the following (secret) criteria:

a) The questions should be balanced so as to ensure a broad range of difficulty.

b) There should be questions referring to all the course chapters, thereby ensuring a homogeneous coverage of the topics studied in the course.

c) the selected questions should stem from all participating students to ensure that each student that had made the effort to prepare MCQs for the game was given a similar ``a priori'' probability to pass the exam. In practice, the instructor took two MCQs from each student, implying that each student had to answer two of her own MCQs in the exam. In order to have a fair evaluation process, one would ideally like to have these MCQs removed from the test. However,  this would have been somewhat impractical, since for each student one would have had to set a slightly different test consisting of 18 MCQs only. Instead, the instructor chose to set the same test for all the students and not to award points for the \emph{first two} correct answers, since he assumed that the students knew the correct answers to their own MCQs (up to a single exception, this assumption was later corroborated by the test results).

In contrast, the exam MCQ test used for the control group during the summer term of the year 2011-2012 consisted of 20 instructor-generated MCQs with topics from all the course chapters. In order to facilitate comparison, no points were awarded for the first two correct answers either (to ensure a fair comparison with the control group, the level of difficulty of 2 questions was chosen to be a very low one). The MCQ tests for both the control group and the experimental group were corrected by requiring that every three wrong answers to MCQs cancel one correct answer. Non-answered questions neither summed nor subtracted points.

For the sake of comparison, the instructor did his best to set a similar level of difficulty for the MCQ tests taken by the experimental and the control group. This was confirmed by a retrospective analysis of the overall quality of the MCQs. The analysis was similar to that given in ref. \onlinecite{Bates2014} and it was based on Anderson and Krathwold's revised version of Bloom's taxonomy \cite{Bloom1956, Anderson2001}. The different taxonomic levels are summarized in Table \ref{tab:taxonomy}, and we assume them to be associated with increasing levels of conceptual difficulty.  Taking the average taxonomic level of the test questions as a criterion, no statistically significant differences between the exam tests for both groups were observed; more specifically, the average taxonomic level of the test questions was slightly higher for the experimental group (3.20) than for the control group (3.05). The instructor also made sure that the overall quality of the distractors in both exam tests was comparable.

\begin{table}[htb]
\caption{\label{tab:taxonomy}
  Categorization levels and explanations for the cognitive domain of Bloom's taxonomy (this table is taken from ref. \onlinecite{Bates2014}).}
\begin{ruledtabular}
\begin{tabular}{cll}
\textrm{Level}&
\textrm{Identifier}&
\textrm{Explanation and interpretation}\\
\colrule
1 & Remember & Factual recall, knowledge, trivial `plugging in' of numbers. \\
2 & Understand & Basic understanding, no calculation necessary. \\
3 & Apply & Implement, calculate or determine. Single topic \\
&& calculation or exercise involving application of knowledge. \\
4 & Analyze & Typically multi-step problem; requires identification \\
&&of problem-solving strategy before executing. \\
5 & Evaluate & Compare and assess various option possibilities; \\
& & often qualitative and conceptual questions. \\
6 & Create & Synthesis of ideas and topics from multiple course \\
& & topics to create significantly challenging problem. \\
\end{tabular}
\end{ruledtabular}
\end{table}

As for the problem-solving part, for both groups of students it consisted of three problems entirely set by the instructor and whose degree of complexity was similar (level 4 in the taxonomic classification of Table \ref{tab:taxonomy}).

\section{\label{sec:results} Results and Discussion}

\subsection{Student satisfaction}

In order to assess the degree of satisfaction of the experimental group with the method, an anonymous was conducted among the 11 students of the experimental group (see Appendix D for an overview of the survey questions).

Seven students indicated that their class attendance was roughly 100 $\%$,  three of them said it was about 75$\%$ and one student said it was about 25$\%$. Even though attendance was not compulsory neither for the control group not for the experimental groups, these figures are realistic and match the instructor's perception. According to the students, the degree of commitment with to the task was very high (8 students said their degree of commitment was 100 $\%$, two said they were committed to $75\%$, an one student said it was $0\%$ (presumably the one having only participated in the game but not having prepared questions for it).

As it turns out, 45 \% of the students had a positive attitude towards the use of the method, whereas the remaining 55\% had a very positive attitude. 73\% think that the method helped them to improve their level of understanding of the course contents. Interestingly, almost 82\% found that the method also helped them to solve problems better. Finally, all the students in the experimental group found the method either useful or very useful (respectively 36\% and 64\%).

Several students made explicit remarks in the sense that the two deadlines set by the instructor to upload questions generated by themselves were very useful to keep up-to-date with the course contents and to prepare for the exam. One student suggested that the rivals in the game be punished by removing wedges for incorrect answers.

\subsection{Statistical analysis}

\begin{figure}
\includegraphics[width=4.5 in]{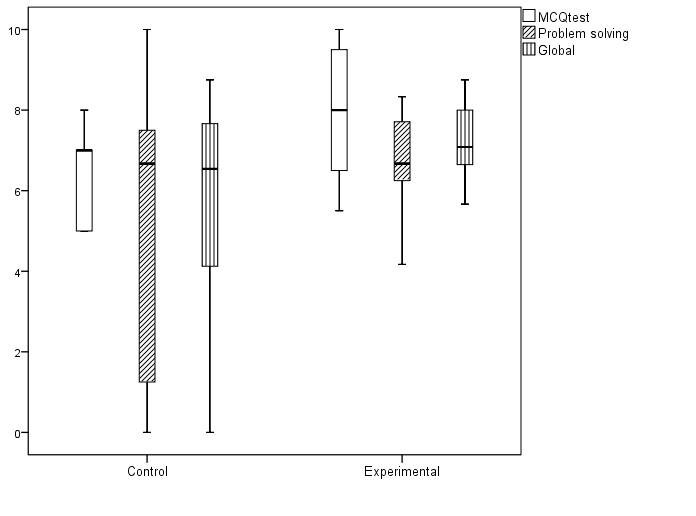}
\caption{\label{fig:1} Box diagram for data distributions.}
\end{figure}

In Table \ref{tab:scores} we show the comparison between the results obtained by the control group and the results obtained by the experimental group in the final exam. Displayed are the mean, the standard deviation (in parentheses) and the median for the MCQ test score, the score in the problem-solving part and the global score. The latter score is obtained as the weighted average of the mean MCQ test score (40\%) and the mean score in problem-solving (60\%).  All scores are normalized to a scale of 10 points (0 being the lowest possible score). In all cases, the mean scores of the experimental group are higher, and the standard deviations are smaller, resulting in a better and less dispersive global performance of the experimental group.  On the other hand, the median for the MCQ test is also higher for the experimental group (albeit the increase is less marked than for the mean score), whereas the medians for problem-solving turn out to be exactly the same. As a result of this, the median of the global score obtained by the experimental group is roughly half a point higher, implying that the best half of the students achieved better results with the experimental method. The box diagram depicted in figure \ref{fig:1} gives a visual summary of the above results.

\begin{table}[htb]
\caption{\label{tab:scores} Mean scores (obtained as averages over the number of students $N$), standard deviations (in parentheses) and median scores.}
\begin{ruledtabular}
\begin{tabular}{lccc}
\textrm{Group}&
\textrm{MCQ test score}&
\textrm{Problem-solving score} & \textrm{Global score} \\
\colrule
Control ($N$=13) & 5.19 (3.07) 7.00 & 5.16 (3.65) 6.67 & 5.18 (3.22) 6.54\\
Experimental ($N$=11) & 7.95 (1.77) 8.00 & 6.59 (1.59) 6.67 & 7.27 (0.99) 7.08 \\
\end{tabular}
\end{ruledtabular}
\end{table}

In order to carry out a more detailed comparison between the score distributions, we have carried out normality tests.  In view of the $p$ values found, we find that normality cannot be assumed for both groups  (see Table \ref{tab:tests}).

\begin{table}[htb]
\caption{\label{tab:tests} Results from normality tests. The asterisks denote lower bounds for significance levels.}
\begin{ruledtabular}
\begin{tabular}{cccccccc}
&& & Kolmogorow-Smirnow && & Shapiro-Wilk & \\
\colrule
& & \textrm{Statistic} & \textrm{df} & \textrm{$p$}  & \textrm{Statistic} & \textrm{df} & \textrm{$p$} \\
\colrule
MCQ test &	Control &	0.260 &	13	& 0.016	&   0.739	& 13 &	0.001 \\
& Experimental &	0.177 &	11 &	 0.200*	  & 0.888 &	11	&  0.131 \\
Problems	& Control & 0.276 &	13 & 0.008	 &  0.847 &	13 &	0.026 \\
& Experimental &	0.247 &	11 &	0.060 & 0.870 & 11	& 0.078 \\
Global	& Control & 0.203 &	13	& 0.148 &  0.827	& 13 &	0.015 \\
& Experimental &	0.146 &	11 &	0.200*	&  0.955	 & 11 & 0.712 \\
\end{tabular}
\end{ruledtabular}
\end{table}

In view of the above, we decided to follow the guidelines of previous works \cite{Gaffney2010, Chasteen2012} and performed a Mann-Whitney U-test \cite{Hollander1999a} (rather than a t-test) in order to draw some conclusions as to whether the observed differences in the scores of the two groups are statistically significant. The obtained results are displayed in Table \ref{tab:mann-whitney}. As can be inferred from the table data, for a significance level equal to 0.05 the null hypothesis must be rejected in the comparison between the MCQ test results for both groups ($p<0.021$), implying that the observed differences for the mean scores are statistically significant. As for the results concerning the score in problem-solving and the global score, even though both are higher for the experimental group, the observed differences do not appear to be statistically significant according to the Mann-Whitney U-test (in both cases one finds $p>0.05$).

\begin{table}[htb]
\caption{\label{tab:mann-whitney} Results from the Mann-Whitney U-test}
\begin{ruledtabular}
\begin{tabular}{lccc}
&
\textrm{MCQ test}&
\textrm{Problem-solving} & \textrm{Global score} \\
\colrule
$p$-value & 0.021 & 0.579 & 0.131\\
\end{tabular}
\end{ruledtabular}
\end{table}

Summarizing, according to our analysis, the experimental group performed significantly better than the control group in the MCQ test. Remarkably enough, we find that the experimental group attained a higher score in problem-solving. While this latter difference in performance does not appear to be statistically significant, it is clear that the time spent by students in designing MCQs was not detrimental to their ability to solve the problems (actually, the possibility of positive feedback for problem solving is not ruled out by the analysis). In global terms, our statistical analysis suggests that enhanced overall performance might be obtained by resorting to games based on student-generated questions.

\section{\label{sec:conc} Conclusions and Outlook}

A game-based method relying on student-generated questions has been used to promote participatory learning and as a knowledge assessment tool. An anonymous survey amongst the students of the experimental group revealed that their degree of satisfaction with the method was high. They pointed out that the method set them in a novel, interesting role calling for abilities they had scarcely used so far, e.g., the rigorous use of scientific language to convey well established facts or ideas of their own in the form of questions and statements (such higher order skills are closely related to key notions stemming from Vygotsky's theory of education, see ref. \onlinecite{Bates2014}). They also pointed out that the use of their own questions in the game and in the final exam definitely invigorated their motivation. They reportedly enjoyed the interaction with their peers when playing the game and had less stress than usual when taking the MCQ test in the final exam.  In conclusion, we can say that was widely acknowledged that the method had contributed to strengthen the students' cognitive and metacognitive skills and thereby to efficiently scaffold their intellectual autonomy.

In his classical essay ``Upon the Aesthetic Education of Man''  the German writer Friedrich Schiller claims that ``humans are only fully human when they play'' \cite{Schiller2000}, thereby setting the stage for the seminal masterpiece ``Homo Ludens" \cite{Huizinga2008} by the renowned cultural theorist Johan Huizinga. This author uses the term ``Play theory'' to define the conceptual space in which play occurs and suggests that the latter is a necessary condition for culture generation. In this context, the term ``culture'' must be understood in a broad sense. As part of the latter, Huizinga argues that epistemology also displays characteristic features of play.

While a quantitative confirmation of Huizinga's standpoint by modern neuroscience is still lacking, we largely share his views on the matter, and it is from this perspective that we devised the pseudoconstructivist bottom-up approach underlying our method.  In our opinion, game-based methods have a considerable pedagogical value, also in the context of higher education. In the sequel of Gardner's theory of multiple intelligences \cite{Gardner1983}, increasing efforts are being devoted to deepen the understanding of the relation between different learning strategies and the activation of neuro-cognitive processes via experiments and computer simulations (see e.g. refs. \onlinecite{Anderson1983} and \onlinecite{Lisle2006} as well as references therein) . In particular, the expected major breakthroughs in this field of research will bring about important implications for the study of the comparative efficiency of game-based methods. In the mean time, valuable complementary insights at a different level can be obtained from studies similar to the present one.

In our specific case, the preparation of MCQs by the students should be viewed as part of the game. In a sense, exam results are ``the prize'' for a good previous performance (students having prepared better for the game are more likely to perform better in the latter as well as in the final exam). While the game can certainly be regarded as a source of entertainment, it would be unfair to leave aside the fact that posing good questions is also an intellectually demanding and highly formative task. Put in simple words, one should bear in mind that ``entertaining'' is the opposite of ``boring'', not the opposite of ``serious''.

Finally, we give a brief summary of the main quantitative results. Our statistical analysis of the exam results reveals a significant improvement of the students' performance in the MCQ test (both the mean score and the median score of the experimental group were higher). We also found that the standard deviation of the MCQ test results was appreciably smaller for the experimental group. This improved performance in the test was not detrimental to the students' ability in problem-solving (the mean was higher for the experimental group, while the median turned out to be the same in both cases).

We plan to extend the present study along different pathways.  For instance, we aim to implement similar methods in other Physics courses with a larger number of students. This requires an intensive dedication of the instructor(s) to review the student-generated questions, hence further human resources will have to be mobilized despite the use of \emph{ad-hoc} web-based tools such as PeerWise.

In view of the above, we strongly believe that game-based methods with student-generated questions can prove useful well beyond the very specific context of undergraduate Physics courses. In this spirit, our hope is that some readers will find the present study inspiring and that this will fuel their motivation to carry out similar experiences in their courses.

\section*{Acknowledgments}

E. A. gratefully acknowledges financial support by the Ministerio de Econom\'{\i}a y Competitividad through Grant No. FIS2013-42840-P (partially financed with FEDER funds) and by the Junta de Extremadura through Grant No. GR10158. P. S. acknowledges financial support by the Vicerrectorado de Docencia y Calidad (Universidad de Extremadura).

\newpage

\section*{Appendix A: Short description of the course program}

Chapter 1 is an introductory chapter aimed at giving a general overview of the basic properties of light and how they were unveiled by various prominent physicists in the course of History.  The laws of reflection and refraction, as well as their derivation in the framework of Newton's corpuscular theory and Huygens' wave theory of light are explained in detail. Young's experiment and the resurrection of the notion of light corpuscles in the framework of the quantum theory of light are also very briefly referred to. The idea is give the students a flavor of important notions that will be revisited and explained in detail in later chapters.

Chapter 2 is devoted to the principles of ray optics and the interaction of light with simple optical systems involving mirrors and lenses. Chapter 3 deals with the basic phenomenology of wave optics (interference and diffraction phenomena, dispersion phenomena, and the properties of polarized light are discussed in detail).   Chapter 4 is devoted to signal transmission and generation in fiber optic systems both from the point of view of ray optics and the theory of electromagnetic modes. Signal distortion due to attenuation and dispersion is comprehensively dealt with, as well as the most common methods to minimize such effects.

Chapter 5 reviews the quantum mechanical principles of light-matter interactions, ranging from Planck's theory for the blackbody spectrum and Einstein's explanation of the photoelectric effect to Bohr's semiclassical model for the hydrogen atom, Heisenberg's uncertainty principle, Schrodinger's equation, the notion of spin and Pauli's principle as well as the selection rules for radiative transitions are subsequently.  Chapter 6 deals with the working principles of lasers and masers and the most common types thereof. Finally, chapter 7 introduces some key notions in semiconductor physics which are subsequently used to explain the physics of photo-emitting diodes (e.g. LEDs and laser diodes) , photodiodes (e.g. PIN or avalanche photodiodes), and photovoltaic cells.

\section*{Appendix B: Examples of original student-generated questions}

The correct answers are marked with a ``(C)''.

\begin{itemize}

\item (Chapter 1) Newton \emph{wrongfully} concluded that light propagates

a) faster through water as it does through air. (C)

b) faster through air as it does through water.

c) with equal velocity through air and through water.

d) with the same velocity in all media.

\item (Chapter 2) Can a converging lens be used as a magnifying glass?

a) Yes, provided that the object's distance to the lens axis lies between $f$ and $2f$.

b) No, never.

c) Yes, provided that the object's distance to the lens axis is $<f$. (C)

d) Yes, provided that the object's distance to the lens axis is $>2f$.

\item (Chapter 3) If the human eye had equal sensitivity in all wavelengths, the sky
would appear to be

a) green.

b) violet. (C)

c) yellow.

d) white.

\item (Chapter 4) What causes modal dispersion in a multimode fiber?

a) Differences in the velocity of each propagation mode along the fiber axis. (C)

b) Destructive interference between propagation modes.

c) The frequency dependence of the mode propagation velocity.

d) None of the above is correct.

\item (Chapter 5) Einstein's theory for the photoelectric effect states that the maximum kinetic energy of the ejected photoelectrons

a) depends on light frequency, light intensity, and on the work function of the photoelectrode material.

b) neither depends on light frequency nor on light intensity, only on the work function.

c) does not depend on light intensity, only on the frequency and the work function. (C)

d) neither depends on light intensity nor on the work function, only on the light frequency.

\item (Chapter 6) In practice, laser light from an optical resonator is not perfectly monochromatic. Broadening may partly arise from

a) a finite bandwith of the auxiliary, short-lived state which feeds the metastable state.

b) a finite bandwith of the metastable state due to dispersion in the atomic velocity at finite temperatures. (C)

c) more than one atomic transition contributing to the laser light.

d) None of the above are correct.

\item  (Chapter 7) Which is the correct $I-V$ relation for a photodiode?

a) $I=I_s\left(e^{eV/kT}-1 \right)-I_L$. (C)

b) $I=I_s\left(e^{eV/kT}-1 \right)$.

c) $I=I_s\left(e^{eV/kT}-1 \right)+I_L$.

d) $I=\left(I_s\,e^{eV/kT}-1 \right)-I_L$.

(The meaning of each symbol had been previously explained in the course).

\end{itemize}

\section*{Appendix C: Trivial Pursuit playing rules}

Even though there are many variants of the game, we focus here on the basic one.  The board consists of a wheel-like track with a central hexagonal hub space and spokes. The track has spaces of different colors, corresponding to different knowledge categories. Each player (or, in our case, each team of players) is given a playing piece. Playing pieces are round and divided into six sections, similar to a pie. A draw sets the starting order of each team. Starting from the central hub, the first team rolls the die and moves their playing piece by the number of spaces indicated by the die. The team must then answer a question related to the category corresponding to the color of the space (some spaces have no color, they rather ask  to roll the die again in order to speed up the game). If the team answers the question correctly, it continues to roll the die and to answer the next question, until a wrong answer is given. Then the turn of the next team comes, and so on. Any time a playing piece lands on the space at the end of a spoke, the team has the chance to fill one of the six sections of the playing piece with a wedge of the color corresponding to the space. As soon as a team fills its playing piece with six wedges of different colors, it has the chance of winning the game by landing its playing piece on the central hub and subsequently providing the correct answer to a question from a category chosen by the other teams.

\section*{Appendix D: Student satisfaction survey form}

1. How many classes of this course did you attend?

a) 0 $\%$   b) 25$\%$   c) 50 $\%$  d) 75 $\%$  e) 100 $\%$

2. How intensively did you commit yourself with the task of the test questions?

a) 0 $\%$   b) 25$\%$   c) 50 $\%$  d) 75 $\%$  e) 100 $\%$

3. Do you think this method has helped you to better understand the course contents?

a) Definitely not b) Probably not c) Maybe d) Probably e) Definitely

4. Do you think this method helped you to better understand and solve the problems set by the instructor?

a) Definitely not b) Probably not c) Maybe d) Probably e) Definitely

5. What is your attitude towards this work methodology?

a) Very negative b) negative c) indifferent d) positive e) very positive

6. Rate the usefulness of the above method

a) not useful at all b) hardly useful c) useful to a certain extent d) rather useful e) very useful

7. Further remarks

\end{document}